\documentstyle[preprint,pra,aps]{revtex}
\begin{document}
%\voffset=-1truein
% \draft command makes pacs numbers print
\draft
\title{A first order phase transition induced by
magnetic field and temperature}
\author{K. J. Singh, S. Chaudhary, 
M. K. Chattopadhyay, M. A. Manekar, S. B. Roy and P. Chaddah}
\address{Low Temperature Physics Laboratory,
Centre for Advanced Technology,\\ Indore 452013, India}
\date{\today}
\maketitle
\begin{abstract}  
Taking the pseudobinary C15-Laves phase compound 
Ce(Fe$_{0.96}$Al$_{0.04}$)$_2$ as a paradigm for studying 
a ferromagnetic(FM) to 
antiferromagnetic(AFM) phase transition, we present interesting thermomagnetic 
history effects in magnetotransport measurements across this FM-AFM transition. We
argue that these distinctive hysteretic features can be used to identify 
the exact nature -first order or second order - of this kind of transition in
magnetic systems where electrical transport is strongly correlated with
the underlying magnetic order. A comparison is made with the similar 
FM-AFM transitions observed in Nd and Pr-based manganese compounds 
with perovskite-type structure.  
\end{abstract}                          
The nature of ferromagnetic (FM) to antiferromagentic (AFM) transition in
the perovskite-type manganese oxide compounds Nd$_{1/2}$Sr$_{1/2}$MnO$_3$
and Pr$_{1/2}$Sr$_{1/2}$MnO$_3$ has been the subject of close scrutiny 
in recent years \cite{1,2}. The FM-AFM transition observed in these 
compounds  is taken as a sort of a prototype of a first order transition, and
certain thermomagnetic features have been highlighted which are thought to be 
generic of a first order phase transition \cite{1,2}. In the same spirit 
we have undertaken a study
of FM-AFM phase transition in the pseudobinary  C15-Laves phase compound 
Ce(Fe$_{0.96}$Al$_{0.04}$)$_2$. 
We find striking thermomagnetic history
effects in magnetotransport measurements across the FM-AFM transition in 
this interesting system. 

CeFe$_2$, with its relatively low Curie temperature ( T$_C\approx$230K) and 
reduced magnetic moment ($\approx$ 2.3$\mu_B/f.u.$) \cite{3}, 
is on the verge of a magnetic instability \cite{4}. 
Neutron measurement has  shown the presence of antiferromagnetic 
fluctuations in the FM ordered state of CeFe$_2$ below 100K \cite{5}.
With small but suitable change in electronic structure caused by doping 
with elements like Co, Al, Ru, Ir, Os and Re at the Fe-site
of CeFe$_2$ \cite{6}, these antiferromagnetic fluctuations get
stabilized into a low temperature AFM state, 
and after certain concentration of 
dopants (usually 5 to 10\%) this AFM phase 
replaces the FM phase altogether \cite{7,8,9,10,11,12,13,14}. 

While most recent experimental efforts are mainly focussed on 
understanding the cause of this magnetic 
instability \cite{15,16} in CeFe$_2$, there exists one other aspect of 
the observed magnetic properties which  needs proper 
attention, viz. the exact nature of the FM-AFM transition. 
We have recently addressed this second question in Ru and Ir-doped 
CeFe$_2$ alloys \cite{17,18}. In this paper we shall 
focus on the Al-doped CeFe$_2$ alloys. 
In contrast to the Ru, Co and Ir doped alloys, a distinct 
co-existence of FM and AFM phase  has been reported
around the FM-AFM transition temperature in Al-doped 
CeFe$_2$ alloys \cite{12}. We report here 
interesting thermomagnetic history dependence of magnetotransport in 
a Ce(Fe$_{0.96}$Al$_{0.04}$)$_2$ alloy. 
We argue that these thermomagnetic history effects
arise due to the first order nature of the FM-AFM transition. 
These effects are broader manifestations of the behavior
reported earlier in manganese compounds \cite{1}, and can be used to
identify a first order FM-AFM transition in a new system, especially in
those where the electrical transport is strongly correlated with the 
underlying magnetic order.

The details of the preparation and characterization of the sample
can be found in Ref.10. 
The samples from the same batch have been used earlier in the study of bulk 
magnetic and transport properties \cite{10}, 
and neutron measurements \cite{12}. We   
have used a superconducting magnet and cryostat system 
(Oxford Instruments, UK) for    
magnetotransport measurements as a  function of temperature 
(T) and applied magnetic field (H). The resistivity is measured
using a standard dc-four probe technique.

The inset of Fig. 1(a) shows the magnetization (M) vs T plot for the
Ce(Fe$_{0.96}$Al$_{0.04}$)$_2$ sample measured with an applied field of 
2 mT. The sample undergoes a paramagnetic (PM) to FM 
transition at around 195K, followed by a lower temperature FM-AFM transition
around 90K. These results are in consonance with the earlier bulk 
properties \cite{10} and neutron measurements \cite{12}. We shall now study 
the field dependence of magnetoresistance in various temperature regimes.
In the main panel of Fig.1 we present resistivity ($\rho$) 
as a function of H at T=3K, 5K, 20K, obtained after initial 
zero-field-cooling (ZFC) the sample to the temperature 
concerned. The $\rho$ vs H plot at T$\geq$120K (not
shown here) is that of a typical ferromagnet, showing clear negative 
magnetoresistance. In the antiferromagnetic regime (see Fig. 1(a)-(c)), we
see the clear signature of a field induced ferromagnetic transition at a field 
H$_M$, where the resistivity decreases sharply with the increase in H. (The
slight increase in $\rho$ in the field regime H$\leq$H$_M$, indicating the
positive magnetoresistance of the AFM state is not quite visible in the 
same scale). Although the change in resistivity due to this field induced 
transition is not as drastic as in Nd$_{1/2}$Sr$_{1/2}$MnO$_3$ \cite{1}, this is 
within an order of magnitude of  those obtained with the similar applied fields 
in Pr$_{1/2}$Sr$_{1/2}$MnO$_3$ \cite{2}. It is to be noted here that 
the FM-AFM transition in the present sample is not accompanied by
a metal-insulator transition and both the FM and AFM states 
remain metallic. On reducing the field from well above
H$_M$, a distinct hysteresis is observed in the $\rho$ vs H plot (see Fig. 
1(a)-(c)). We attribute this to the first order nature of the field induced
AFM-FM transition. While reducing H from
well inside the FM state i.e. H$>>H_M$, the FM state  continues to exist as 
supercooled metastable state below H$_M$ up to a certain     
metastability field H$^*$ \cite{19}. Between H$_M$ 
and H$^*$ fluctuations will help in the formation 
of droplets of the stable AFM state, and at H$^*$ 
an infinitisimal fluctuation will 
drive the whole system to the stable AFM state. Similar hysteresis in the 
$\rho$-H plots of  Nd$_{1/2}$Sr$_{1/2}$MnO$_3$ and 
Pr$_{1/2}$Sr$_{1/2}$MnO$_3$ has also been attributed to the first 
order nature of the phase transition \cite{1,2}. 
The role of thermal fluctuations is expected to be
reduced in the very low temperature regime, and this 
is clearly seen in the $\rho$-H plots at T=3K and 5K (see Fig. 1(b) and (c)). In 
this temperature regime on reduction of the applied H to zero  the $\rho$(H=0) lies
distinctly below the intial ZFC-$\rho$(H=0), thus giving rise to an open
hysteresis loop. This kind of open hysteresis loop has earlier been 
reported for Nd$_{1/2}$Sr$_{1/2}$MnO$_3$ \cite{1} but not for 
Pr$_{1/2}$Sr$_{1/2}$MnO$_3$ \cite{2}. We attribute this behaviour to the 
existence of a residual metastable FM state even when the applied H is reduced 
to zero. On increasing H on the negative side, the $\rho$(H) curve 
is clearly not symmetric to the virgin $\rho$(H) curve on
the positive H side (see Fig. 1(b)-(c)).
The field induced AFM-FM transition, however, takes place at the same $|H_M|$ 
(see Fig. 1(b)-(c)). A distinct hysteresis is observed 
on reducing H from the
same $|H_{max}|$ as in the positive side.
The envelope hysteresis loop now closes at H=0, i.e. it 
merges with the starting value where from the field excursion on the negative
direction 
had started. Further on increasing  H to the positive side, the $\rho$(H) 
curve now follows a path which is distinctly below the virgin $\rho$(H) curve
but very symmetric to the $\rho$(H) curve on the negative side in the 
increasing H cycle. This $\rho$(H) curve (henceforth will be termed as
forward envelope curve) merges with the virgin curve in the H-regime well
beyond H$_M$. We have checked this distinct difference between 
the virgin curve and the forward envelope curve in the negative H side also, 
by drawing a virgin curve in the negative H direction after zero field 
cooling the sample (see dashed line in Fig. 1(b)-(c)). This
anomalous behaviour of virgin curve lying distinctly outside the envelope
curve is observed \cite{20} in the field dependence of 
magnetization as well (see inset of Fig. 1(b)).

From a closer inspection of the published results on 
the Nd$_{1/2}$Sr$_{1/2}$MnO$_3$ sample \cite{1},  
we expect similar behaviour to take place in that sample in the 
temperature regime T$\leq$20K. Specifically the $\rho$-H curve was shown to
have an open hysteresis-loop (see figs. 2 B to 2D of ref.1). We expect that
if H was reduced to -12T in that sample, and then raised back to +12T, this
second leg would be closed similar to our Fig. 1(a). Similarly an isothermal
M-H measurement in that compound should show a virgin curve lying outside the
envelope curve as in the inset of Fig. 1(b).

Another interesting aspect worth
noting in our present sample is that the slope of the $\rho$(H) return envelope 
curves change sharply on crossing H=0 in either direction 
(see Fig. 1(b)-(c)). 
It almost flattens in the low field regime H$<|H_M|$ 
on either side of H=0, as if the domains of the residual 
FM state still retain their previous memory.   
With the striking similarity of the field dependence of resistivity in
Nd$_{1/2}$Sr$_{1/2}$MnO$_3$ (see Fig. 2(b)-(d) of Ref.1)
 with those in the positive H
cycle in the present sample (Fig. 1(b)-(c)), it is quite tempting to predict
the similar sharp change in slope of R(H), as H changes sign of H in
Nd$_{1/2}$Sr$_{1/2}$MnO$_3$ as well.

Supercooling/superheating and metastability have been identified as
key elements to explain the thermomagnetic history effects associated 
with the first order FM-AFM transition in perovskite-type 
manganese oxide systems \cite{1}. 
To explain the strong temperature dependence
of this thermomagnetic irreversibility in our present system, we now invoke
in addition the concept of the limit of the metastability H$^*$(T) ( or
T$^*$(H)). The implicit assumption involved here is 
that the difference between 
the phase transition line H$_M$(T) ( or T$_N$(H)) and the limit of 
metastability H$^*$(T) ( or T$^*$(H) ) widens with the decrease (increase)
in temperature (magnetic field). To support this assumption 
we shall now study the temperature
dependence of resistivity in the presence of various applied magnetic fields.

In Fig. 2 we present $\rho$ vs T plots with H=0, 0.5T, 2T and 3T. 
Appearance of magnetic superzones \cite{10,11} at the FM-AFM transition (T$_N$) give rise
to a distinct structure in the form of a local minimum in the $\rho$(T) (see
Fig.2). There is a marked hysteresis associated with this transition, 
emphasizing the first order nature of the transition. We argue that the 
FM (AFM) phase exists as supercooled (superheated) 
metastable phase in the cooling (heating) cycle 
in this hysteretic temperature regime. In the high (low) temperature 
reversible regime the only magnetic phase is the stable FM (AFM) phase.
In an earlier zero field 
neutron measurement \cite{12} a clear co-existence 
of FM-AFM phase was observed
for a substantial temperature regime below the onset of phase transtion. This
alongwith the gradual development of the cubic to rhombohedral 
structual distortion \cite{12} as the temperature is lowered, 
fit naturally in our present picture of first order
phase transition. In the presence of finite H \cite{21} the T$_N$ is suppressed and the
hysteresis is enhanced substantially, so much so that with H=2T no
reversible regime is observed below T$_N$
down to 5K. This in turn implies that the limit
of metastability T$^*$(H) (below which one should see the reversible response
of the stable AFM phase) gets suppressed even faster in 
comparison to T$_N$(H), and the metastable regime (encompassed between T$_N$(H)
and T$^*$(H)) widens with the decrease  in T or with increase in H. 
This is again in perfect consonance with our observation 
in the isothermal field dependence of
resistivity.

Summarizing our results, the important findings of the present study 
are the following:
\begin{enumerate}
\item The butterfly $\rho$(H) hysteresis loop with anomalous virgin curve, 
which does not close at H=0 in the low temperature regime. This is
complimented with magnetization study as well.
\item The distinct hysteresis in the $\rho$(T) curve at AFM-FM transition,
which gets enhanced in the presence of applied magnetic field.
\end{enumerate}      
These observations along with the earlier neutron measurements \cite{12} 
help to establish the first order nature of the AFM-FM transition in
the CeFe$_2$-based pseudobinary systems. This information in turn 
will be important for developing a theoretical model to explain 
the interesting electro-magnetic properties of CeFe$_2$, which does not
exist so far. Most importantly, thermomagnetic history effects 
observed here can be used as generic features to identify a first order
FM-AFM phase transition. 
In fact the  double hysteresis or butterfly loop in
polarization  measurements  is regularly used to identify a first order 
transition in the ferroelectric/antiferroelectric materials \cite{22}. 
(However, we are unaware of any report regarding the anomalous 
nature of the virgin curve associated with the butterfly loops
in the ferroelectric materials). Such a field induced
first order transition can be explained in terms of free energy curves 
obtained by expanding in a power series in polarization  and retaining 
only terms with even powers up to sixth order \cite{22}. Phase coexistence 
and metastability across the first order phase transition, and hence hysteresis
arise naturally out of such free energy curves. 
Although, sharp rise in field induced magnetization 
(and associated hysteresis) \cite{13}
and sharp drop in the field induced resistivity \cite{14,23} 
in magnetic systems are regularly
taken as  signatures of AFM-FM transition, no clearcut inference can be
made from such observations regarding the exact nature of the phase transition. 
 Our present study along with those on perovskite-type
manganese oxide systems \cite{1,2} attempt to 
fill-up this gap and provide some relatively easy means to identify
the nature of such magnetic transitions from standard bulk properties 
measurements. It should be noted here that the resistivity/magnetoresistance
is not a thermodynamic quantity, whereas magnetization/staggered magnetization
is, and can be used as an order parameter in the Landau type free energy 
expansion. However, intricate correlation between the resistivity/magnetoresistance
and magnetization/staggered magnetization across the FM-AFM transition, makes
these magnetotransport properties suitable observables in the study of the
nature of the phase transition. Further, these samples with well characterized  
electro-magnetic properties can be used as
paradigms to study the various interesting aspects of a first
order transition, namely nucleation and growth, and path dependence 
of the transition, in a relatively easy and reproducible manner. 
Such studies are not that easy and reproducible in the more 
common cases of first order 
transition, like melting (solidification) of solids (liquids) or vaporization
(condensation) of liquids (gases).

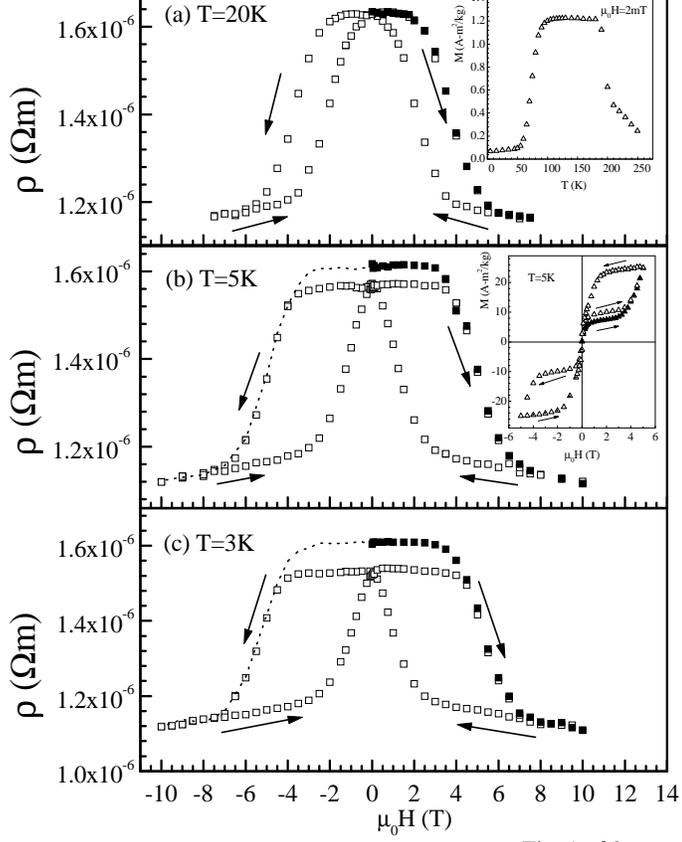
\begin{figure}
\caption{Resistivity vs field plots of Ce(Fe$_{0.96}$Al$_{0.04}$)$_2$ 
at T=20K, 5K and 3K. The open square symbols represent the envelope
curves initiated between the respective $\pm \mu_0 H_{max}$ as
indicated in the figures. Closed squares show the "virgin" curve 
where the sample is cooled in zero field and the field is then
raised to 10T. Inset of Fig. 1(a) shows the magnetization vs temperature
plot obtained with a field of 2 mT. Inset of Fig. 1(b) shows the 
magnetization vs field plot at T=5K. The virgin (envelope) curve is
shown by filled (open) triangle symbols; note that the virgin curve is lying
outside the envelope curve.}
\end{figure} 
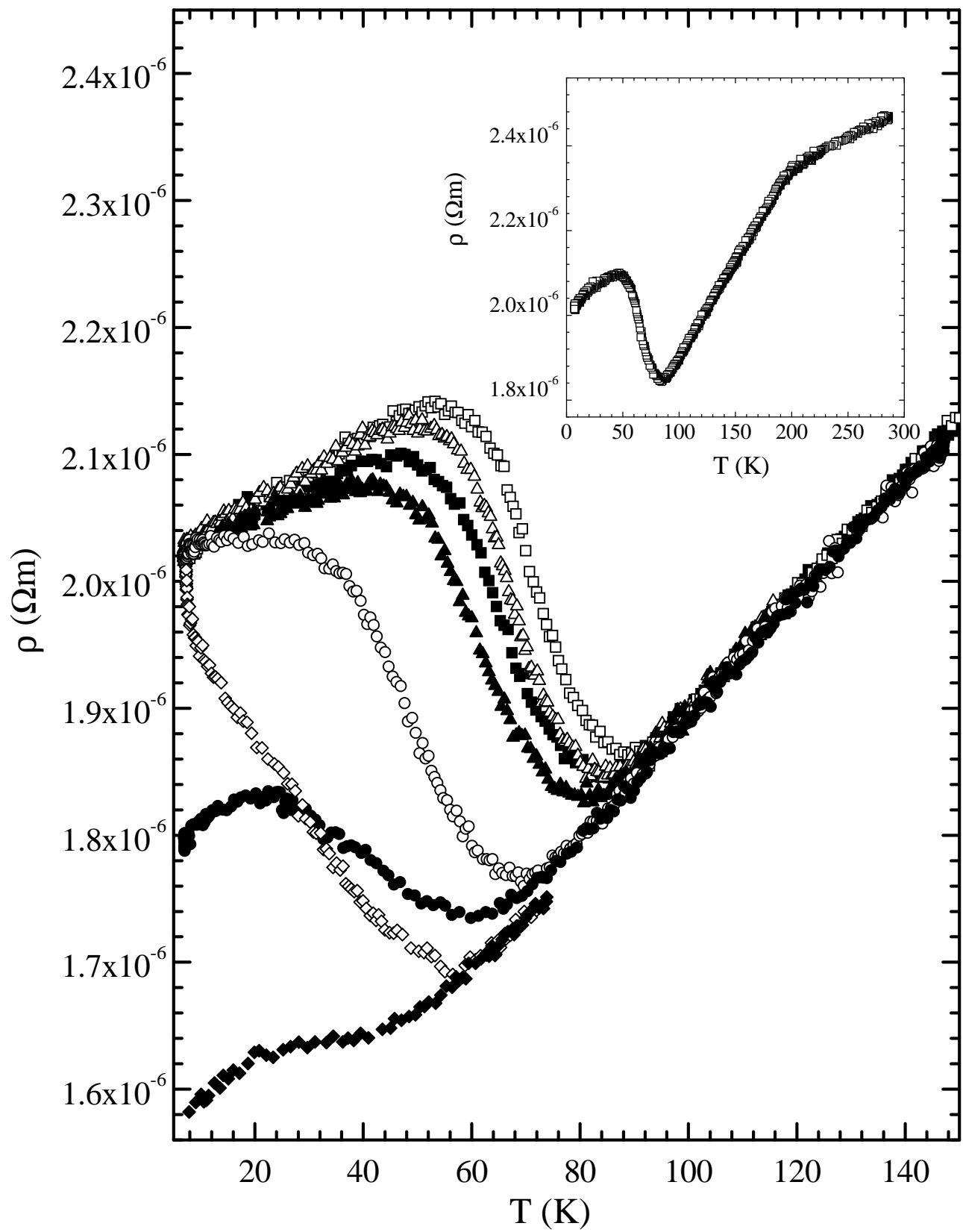
\begin{figure}
\caption{Resistivity ($\rho$) vs temperature plots in the presence of various applied 
fields H=0, 0.5T, 2T and 3T shown by square, triangle, circle and diamond 
symbols respectively. The experimental protocol is described in Ref.21. The
open (filled) symbols show the $\rho$(H)-T behaviour of the sample recorded during
warming (cooling). Inset shows the zero field $\rho$-T plot showing both 
the PM-FM and FM-AFM transitions.}
\end{figure}

\begin{references}
\bibitem{21}In the resistivity measurements in presence of finite H, we first zero-field
cool the sample to the lowest temperature of measurement, switch on the
field, warm up the sample to the highest temperature of measurement, and then
cool back the sample (in presence of the field) across the phase 
transition temperature down to the lowest temperature of measurement.
\end{references}
\end{document}